%% file: main.tex
\documentclass[letterpaper,twocolumn,10pt]{article}
\PassOptionsToPackage{hyphens}{url}
\usepackage{usenix}

\usepackage{tikz}
\usepackage{amsmath}
\usepackage{booktabs}
\usepackage{enumitem}
\usepackage{comment}
\usepackage{subcaption}

\usepackage{siunitx}
\DeclareSIUnit\sample{S}
\DeclareSIUnit[group-minimum-digits=3]\usd{USD}
\DeclareSIUnit[quantity-product = ]\percent{\char`\%}
\newcommand{\mathqty}[2][]{\qty[parse-numbers=false,#1]{#2}}

\newcommand{\sysname}[0]{\mbox{\textsc{SatIQ}}}

\usepackage{pifont}

\newcommand{\autobox}[1]{%
    \centering
    \resizebox{%
        \ifdim\width>\linewidth
            \linewidth
        \else
            \width
        \fi
    }{!}{#1}%
}

\begin{document}

\date{}

\title{\Large \bf SATversary: Adversarial Attacks and Defenses for Satellite Fingerprinting}

\author{
{\rm Joshua Smailes}\\
University of Oxford
\and
{\rm Sebastian K\"{o}hler}\\
University of Oxford
\and
{\rm Simon Birnbach}\\
University of Oxford
\and
{\rm Martin Strohmeier}\\
armasuisse Science + Technology
\and
{\rm Ivan Martinovic}\\
University of Oxford
}

\maketitle

\input{abstract.tex}

\input{motivation.tex}

\input{background.tex}
\input{threat-model.tex}

\input{experiment-design.tex}
\input{results.tex}

\input{discussion.tex}

\input{conclusion.tex}

\section*{Acknowledgments}
We would like to thank the System Security Group at ETH Z\"{u}rich for allowing us to use their hardware for the channel model experiments, and Harshad Sathaye in particular for his help and insights while setting up the hardware.
We would also like to thank armasuisse Science + Technology for their support during this work.
Joshua was supported by the Engineering and Physical Sciences Research Council (EPSRC).
Sebastian was supported by the Royal Academy of Engineering and the Office of the Chief Science Adviser for National Security under the UK Intelligence Community Postdoctoral Research Fellowships programme.
Simon was supported by the Government Office for Science and the Royal Academy of Engineering under the UK Intelligence Community Postdoctoral Research Fellowships scheme.

\appendix

\cleardoublepage
\bibliographystyle{plainurl}
\bibliography{main}

\cleardoublepage
\input{appendices/experiment-hardware.tex}

\input{appendices/gan-layers.tex}
\input{appendices/extended-results.tex}

\end{document}

%% file: abstract.tex
\begin{abstract}

Due to the increasing threat of attacks on satellite systems, novel countermeasures have been developed to provide additional security.
Among these, there has been a particular interest in transmitter fingerprinting, which authenticates transmitters by looking at characteristics expressed in the physical layer signal.
These systems rely heavily upon statistical methods and machine learning, and are therefore vulnerable to a range of attacks.
The severity of this threat in a fingerprinting context is currently not well understood.

In this paper we evaluate a range of attacks against satellite fingerprinting, building on previous works by looking at attacks optimized to target the fingerprinting system for maximal impact.
We design optimized jamming, dataset poisoning, and spoofing attacks, evaluating them in the real world against the \sysname{} fingerprinting system designed to authenticate Iridium transmitters, and using a wireless channel emulator to achieve realistic channel conditions.
We show that an optimized jamming signal can cause a \qty{50}{\percent} error rate with attacker-to-victim ratios as low as~\qty{-30}{\decibel} (far less power than traditional jamming techniques), and demonstrate successful spoofing attacks, with an attacker successfully removing their own transmitter's fingerprint from messages.
We also present a viable dataset poisoning attack, enabling persistent message spoofing by altering stored data to include the fingerprint of the attacker's transmitter.

Finally, we show that a model trained to optimize spoofing attacks can also be used to detect spoofing and replay attacks, even when it has never seen the attacker's transmitter before.
This technique works even when the training dataset includes only a single transmitter, enabling fingerprinting to be used to protect small constellations and even individual satellites, providing additional protection where it is needed the most.

\end{abstract}

%% file: motivation.tex
\section{Motivation}\label{sec:motivation}

As the rise of off-the-shelf Software Defined Radio (SDR) hardware makes it easier for even hobbyist-level actors to attack radio signals, legacy satellite systems have been made particularly vulnerable due to their lack of cryptographic security~\cite{manulisCyber2020,fritzSpace}.
In response, physical layer fingerprinting has emerged as a promising countermeasure: by looking at impairments on the physical layer radio signal, transmitters can be characterized and differentiated from one another.
Crucially, this can separate legitimate transmitters from attacker-controlled SDRs, providing robust authentication even in the absence of cryptography.
At its face these systems grant enhanced security, but it is not known to what extent they are vulnerable against direct attacks on the fingerprinting system itself -- existing evaluations of attacks are limited to simple replay and jamming~\cite{rehmanAnalysis2014,edmanActive2009,irfanPreventing2025}, or make use of Arbitrary Waveform Generator (AWG) hardware, which is prohibitively expensive for many attackers~\cite{danevAttacks2010}.

In this paper we provide the first end-to-end evaluation of wireless fingerprinting under optimized jamming, spoofing, and poisoning attacks, assessing the extent to which these systems are vulnerable to attackers equipped with off-the-shelf hardware.
We focus on satellite systems as a case study, due to the particular relevance of fingerprinting in this area -- legacy satellites are uniquely both vulnerable to attacks on the wireless channel and prohibitively expensive to upgrade or replace, making fingerprinting a particularly appealing countermeasure.
In undertaking this work, we can better understand the threats associated with fingerprint-based authentication alongside its benefits, enabling operators to make well-informed decisions surrounding its implementation.

Our contributions are as follows:
\begin{itemize}
    \item We demonstrate optimized jamming attacks against satellite fingerprinting, showing that it is possible to cause significant disruption at very low amplitudes.
    \item We show that attackers can carry out poisoning attacks, modifying reference messages over time through incremental updates to cause their own transmitter to be accepted as legitimate.
    \item We evaluate optimized spoofing attacks, showing that attackers can partially remove the fingerprint of their own transmitting hardware when replaying messages to bypass fingerprint authentication.
    \item We demonstrate the use of Generative Adversarial Networks (GANs) as a countermeasure to attacks, showing that the discriminator from a GAN trained to carry out spoofing attacks can be used to detect attacks with high accuracy, even from previously unseen transmitters.
\end{itemize}

The implications of these findings are significant, highlighting the potential for fingerprinting systems to be vulnerable to targeted attacks, and the importance of using other countermeasures alongside fingerprinting.
Furthermore, the GAN countermeasure also has the benefit of being possible in systems with only a single transmitter -- contrary to previous systems which require a large dataset of many different transmitters, a model can be trained on data from only a single transmitter to differentiate between legitimate and malicious communication.
This will enable operators of small constellations or single satellites to gain additional protection against attacks by deploying this system, potentially alongside other countermeasures.

%% file: background.tex
\section{Background}\label{sec:background}

In this section we introduce fingerprinting as a concept, looking in particular at its use in a satellite context and the difficulty of deployment in this area.
We also introduce GANs and dataset poisoning attacks.
Finally, we explore related work, looking in particular at adversarial attacks on fingerprinting, and similar attacks in the context of biometric systems.

\subsection{Fingerprinting for Authentication}

Wireless fingerprinting enables the authentication of transmitters by extracting characteristics of the transmitter as expressed in the physical layer signal.
These can be compared to previous messages to verify that they match, and that the message was therefore sent by the legitimate transmitter rather than spoofed or replayed by an SDR-equipped attacker.

There is already a large base of existing research in radio transmitter fingerprinting~\cite{soltaniehReview2020}, with techniques largely falling into either transient fingerprinting (looking at the start of the signal)~\cite{rasmussenImplications2007} and steady-state fingerprinting (looking at the modulated data portion)~\cite{foruhandehSpotr2020}.
Features can be extracted by a range of techniques: combining low sample rate messages~\cite{oligeriPASTAI2020,wangRadio2022}, looking at high frequency components~\cite{basseyIntrusion2019}, extracting from the frequency domain~\cite{kennedyRadio2008}, and more.
These have been applied to a number of terrestrial systems including RFID~\cite{bertonciniWavelet2011}, the ADS-B air traffic control system~\cite{birnbachAdaptable2023}, Bluetooth~\cite{barbeauDetection2006}, and WiFi~\cite{suskiiiUsing2008}.

The deployment of fingerprints as a post-hoc means of authentication is particularly useful in satellite systems, due to their inherent reliance on wireless communication and the immense difficulty of repairs and upgrades once deployed.
In this environment, fingerprints can be used to authenticate downlinked messages on the ground without making any changes to the space segment.
However, it is also more difficult to extract fingerprints from satellite communication compared to other ground-based Radio Frequency (RF) systems due to the high degree of atmospheric distortion and attenuation, masking transmitter characteristics, and requiring specialized techniques to overcome.
This has been explored by some recent works, looking in particular at GPS navigation satellites~\cite{foruhandehSpotr2020} and the Iridium constellation~\cite{oligeriPASTAI2020,smailesWatch2023,smailesSatIQ2025}.
Among these, \sysname{}~\cite{smailesWatch2023} provides a helpful experimental foundation, since the code, dataset, and model have been made available, and are easy to adapt to an adversarial context.
Alongside an increased interest from an academic perspective, fingerprinting has also gained the attention of the commercial and public sectors: the European Space Agency (ESA) have recently allocated funding to explore applications for fingerprinting in the satellite context~\cite{esaApplications2025}, and in 2021 the company Expedition Technology were awarded a contract by the US DARPA to expand their RF fingerprinting and spectrum characterization capabilities~\cite{expeditionTechnology2021}.

\subsection{Generative Adversarial Networks}

Generative Adversarial Networks (GANs) are a popular technique for image generation~\cite{baoCVAEGAN2017}, and can be adapted to work in other contexts.
The architecture is characterized by the combination of two components: a \textit{generator}, which creates adversarial data, and a \textit{discriminator}, which must tell the difference between legitimate and generated data~\cite{mirzaConditional2014}.
Both of these are trained at the same time so that, over time, the discriminator gets better at distinguishing falsified and real data, thus forcing the generator to get better at synthesizing realistic data.
We adapt this architecture to work with radio signals, training a GAN to generate signals that remove the fingerprint of an attacker-controlled SDR.

\subsection{Poisoning Attacks}

Data poisoning attacks involve the introduction of malicious data into a system, affecting its operation in a specific way~\cite{tianComprehensive2022,wangThreats2022,barrenoCan2006,lovisottoBiometric2020}.
These attacks can take place during the training phase, by adding false data into the training dataset in order to reduce performance, increase misclassifications, or falsely accept specific inputs.
Alternatively, they can occur during normal operation, altering the ground truth over time through the gradual introduction of adversarial examples -- this is particularly effective in biometric systems, in which inputs are compared against previous data~\cite{biggioAdversarial2015}.
The presence of update mechanisms (which are often triggered by changes to the input) enables the attacker to gradually shift the ground truth against which new inputs are compared, such that a given target input is accepted.

Poisoning techniques are broadly compatible with the satellite fingerprinting context, although to the best of our knowledge no prior works specifically target satellite fingerprinting systems at the time of writing.
We demonstrate poisoning attacks on satellite fingerprinting in Section~\ref{sec:experiment-design-poisoning}.

\subsection{Related Work}\label{sec:related-work}

The authors of~\cite{adesinaAdversarial2022} provide an overview of adversarial attacks on RF machine learning systems.
This includes attacks on signal and modulation classification systems~\cite{kimOvertheAir2020,kimChannelAware2021,kokalj-filipovicTargeted2019}, jamming attacks~\cite{shiAdversarial2018}, and attacks on RF fingerprinting systems~\cite{restucciaGeneralized2020,karunaratnePenetrating2021}.
In~\cite{restucciaGeneralized2020} a generalized approach is given for adversarial jamming and spoofing against a target neural network, and the authors demonstrate its effectiveness against a classifier for the ``ADS-B'' protocol used in aviation, and a signal modulation classifier.
We use a similar approach in this paper, using gradient descent to find optimized jamming signals.
In~\cite{karunaratnePenetrating2021}, the authors instead use reinforcement learning techniques to fool a discriminator classifier into accepting messages from an attacker-controlled transmitter, based on binary classification alone, evaluating the approach using 8 SDRs.
The desired outcome in this case is similar to our spoofing experiments, but we examine a much larger set of transmitters and focus on real-world systems.
Finally, in~\cite{allaRobust2024} the authors use SDRs to gather a dataset for fingerprinting, but all imperfections are introduced at the software level, and the impact of the SDRs is not measured.

There has also been some work looking at direct impersonation of device fingerprints: the authors of~\cite{danevAttacks2010} replay messages using an AWG, resulting in the duplication of transmitter fingerprints.
In~\cite{rehmanAnalysis2014} a similar approach is attempted using SDR hardware at lower sample rates, with limited success.
We build upon these works, using SDRs to carry out attacks without perfectly impersonating the transmitter.

Many of the attacks explored in related works are against classifier-based systems, with a particular focus on image classification~\cite{machadoAdversarial2021}.
In the case of classifiers, attacks naturally target misclassifications within the system.
However, when targeting an embedding-based system like \sysname{}, attacks must instead focus on the distance metric produced by the model, aiming to either increase the distance between legitimate messages and their stored examples to jam communication through false rejections, or spoof communication by decreasing this distance for replayed messages.
This aligns more closely with adversarial research in biometric systems, in which there already exists a good amount of work.
For instance, in~\cite{marroneAdversarial2019} the authors use gradient descent to produce adversarial perturbations for misclassification in face recognition systems, and in~\cite{martinez-diazevaluation2011} the authors use hill climbing (a similar approach) to find optimized attacks on a human fingerprint recognition system.
Finally, in~\cite{gomez-alanisGANBA2022} the authors train a GAN to perform spoofing attacks on voice recognition systems, and discuss its usefulness as a countermeasure to these attacks.
We explore similar approaches to each of these, adapting known methodology to work within the high degree of noise and attenuation of satellite communication, and designing new model architectures which work well despite these limitations.

%% file: threat-model.tex
\section{Threat Model}\label{sec:threat-model}

In this paper, we consider attacks against a satellite ground system in which downlinked communication from satellites is protected by fingerprint-based authentication.
Therefore, any attacks against this system must first overcome the fingerprinting defenses.
Depending on the attacker goal, this may be achieved either by denying service through increased false rejections, or by altering the identity of an attacker-controlled transmitter to cause its messages to be falsely accepted.

\subsection{Attack Strategies}

We look in particular at three attack strategies:

\subsubsection*{Optimized Jamming (Fingerprint Disruption)}
The attacker crafts a jamming signal optimized to disrupt the fingerprint as much as possible within their power limitations, disrupting the availability of the system by altering the fingerprint of legitimate communication.
This builds on prior works, which evaluate jamming attacks on fingerprinting systems using standard jamming techniques~\cite{irfanPreventing2025,smailesSticky2024}.
Targeting the jamming signal against the fingerprinting model itself increases its effectiveness, enabling the attacker to use less power to achieve the same level of disruption.

\subsubsection*{Fingerprint Data Poisoning}
The attacker modifies the stored example messages which are used to identify a given transmitter, gradually altering them to resemble their own hardware -- this causes their transmitter's fingerprint to be accepted as legitimate.
Depending on the poisoning process, the result can be inclusive (the victim's original hardware is accepted alongside the attacker's transmitter) or exclusive (the victim hardware is rejected, and only the attacker is accepted).

\subsubsection*{Optimized Spoofing (Fingerprint Masking)}
The attacker wishes to transmit an arbitrary message from a ground-based SDR and have it received as legitimate at the victim receiver.
This is achieved by preceding their message with a replayed or generated message header mimicking that of the legitimate satellite, in an attempt to replicate the fingerprint.
However, the fingerprint is altered by their transmitter hardware, so they must undo the effect of their transmitter's fingerprint in order for the message to be accepted.

\subsection{Capabilities}

We assume the attacker has access to commercial off-the-shelf SDR hardware, an appropriate amplifier, and an antenna, allowing them to transmit signals within the vicinity of the victim ground station.
The attacker is assumed to run their SDR at a sample rate matching the target fingerprinting system -- \qty{25}{\mega\sample/\second} in our experiments.
This can be achieved even using low-cost SDR hardware such as the ``LimeSDR Mini 2.0'', costing only \qty{440}{\usd}.%
\footnote{Price accurate as of February 2026.}

Since the attacker transmits messages over the air, we assume the attacker can achieve time synchronization with the victim receiver, and can therefore transmit targeted interference (for example, jamming signals) over the top of legitimate messages, or send their own messages to be picked up by the victim receiver.
However, they cannot achieve perfect synchronization at the symbol or phase level -- to do so would require a feedback loop with the victim ground system, which is usually not feasible for attackers.
When messages are transmitted, the signal is impaired by the fingerprint of the attacker's transmitter, and is further affected by multipath fading and other characteristics of the physical channel.
These must be counteracted in the case of spoofing attacks, making them significantly more challenging to execute~\cite{rehmanAnalysis2014}.%
\footnote{It has already been established in related work that an attacker equipped with an AWG can perfectly replicate signals down to the fingerprint in an experimental setting~\cite{danevAttacks2010}, but due to the high cost of this hardware we consider it to be out of scope for this work.}

During poisoning attacks, we assume the attacker has some mechanism by which they can introduce malicious data into the fingerprinting system.
This could be achieved via spoofing, or using an alternative side channel into the system itself (for example, if the machine hosting the fingerprint examples has been briefly compromised).
Although this attack is initially more challenging to execute, it has a much higher success rate in the long term: once the data has been altered to match the attacker's hardware, they do not need to worry about masking their own fingerprint, as it is being accepted with the same rate as any other transmitter.
If inclusive poisoning is used, the victim's original transmitter will also continue to be accepted alongside the attacker's hardware, increasing the longevity of the attack.

Finally, for the jamming and poisoning experiments we assume the attacker to have access to the underlying weights of the fingerprinting model.
This is a reasonable assumption to make in many cases, since for many existing fingerprinting systems the code, dataset, and model weights are openly available.
However, even if the attacker does not have access to this data, it has been shown in previous works that the same attacks can be achieved by training a ``surrogate model'' on a similar dataset, executing the attacks on this model, and transferring the attack to the original system~\cite{demontisWhy2019,lovisottoBiometric2020}.
We demonstrate this through the use of a GAN in the ``fingerprint masking'' attack.

%% file: experiment-design.tex
\section{Experiment Design}\label{sec:experiment-design}

We next design experiments to evaluate each of the attacks described in the previous section.
We look at three classes of attack: jamming, poisoning, and spoofing.
All code, datasets, and trained models used in this paper will be made freely and openly available on publication.

\subsection{Configuration}

We start by describing the software and hardware configuration common to each of the experiments.

\subsubsection{Dataset and Software Configuration}

For each of our experiments, we use a trained \sysname{} model provided by the authors of~\cite{smailesWatch2023}.
This model takes message headers from Iridium transmitters at a sample rate of \qty{25}{\mega\sample/\second} and condenses them into a fingerprint, which is then compared to reference examples for the same transmitter, resulting in a distance metric which is used to authenticate the transmitter.

We also use the publicly available dataset from~\cite{smailesWatch2023}, so that the data is compatible with the model and the results from this paper can be easily reproduced.
This data comprises \num{1.7}~million Iridium messages collected over \num{40}~days, with data from \num{66}~satellites, each of which have \num{48}~transmitters~\cite{veenemanIridium2021}.
In all the experiments in this paper, all signals are generated at the same sample rate as the original dataset, \qty{25}{\mega\sample/\second}.
They are then scaled to a consistent overall energy level, and the signal is rotated by a random phase offset between $0$ and $2\pi$, to mimic the effect of an attacker whose signal is not phase synchronized with the victim.%
\footnote{Note that only the attacker's modifications are rotated in this manner -- the original dataset corrects for phase offset, so the victim messages are not rotated.}

\subsubsection{Hardware Configuration}\label{sec:hardware-configuration}

\begin{figure}
    \centering\includegraphics[width=\linewidth]{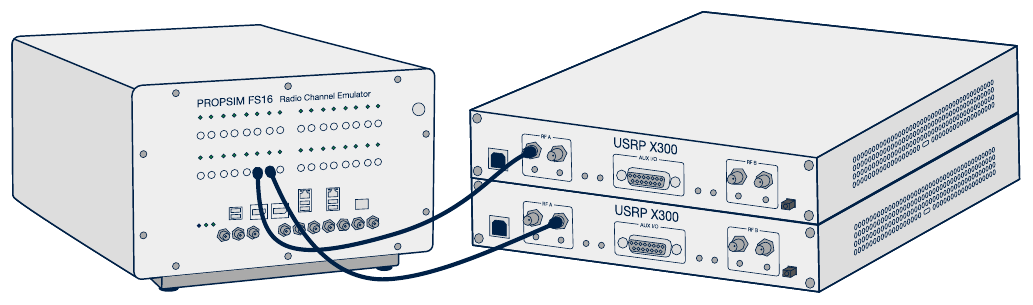}
    \caption{The transmit-receive loop used during experiments to measure the impact of the attacker's transmitter and channel effects on the fingerprint. Messages are transmitted by one SDR, pass over a wire and through the PROPSIM channel emulator, before being received by the other SDR.}
    \label{fig:experiment-design-propsim}
\end{figure}

When carrying out jamming or spoofing attacks, the attacker's own transmitter fingerprint is crucial to the attack's success or failure.
These experiments must therefore include a hardware component.
We achieve this by building a transmit-receive loop composed of two SDRs, illustrated in Figure~\ref{fig:experiment-design-propsim} -- messages are sent to one SDR, transmitted over the cable, and received at the other end.
This leaves the contents of the message untouched, but the physical layer characteristics altered by the SDRs' fingerprints, which must be counteracted by the attacker during spoofing attacks.

To ensure the physical layer is as accurate as possible, a PROPSIM wireless channel emulator is also included in the hardware loop.
This device simulates wireless channel conditions to provide realistic attenuation and multipath fading without transmitting signals over the air, avoiding the risk of violating broadcast regulations and providing a high degree of control over channel conditions.
The emulator provides implementations of the 3GPP standard channel models for non-terrestrial networks, ensuring our experiments accurately reflect real-world channel characteristics~\cite{3gppTechnical2020}.
We use three configurations of the channel model to cover a broad range of channel conditions, representing the best case for attackers with a clean channel, the worst case with high multipathing and no line of sight, and a purely wired channel as a control.
These are configured as follows:
\begin{itemize}
    \item \textbf{Best Case:} ``Rural'' user environment, tapped delay line model ``D'' (line of sight).
    \item \textbf{Worst Case:} ``Dense Urban'' user environment, tapped delay line with model ``A'' (no line of sight).
    \item \textbf{Wired Channel:} The channel emulator is not used, and the SDRs are connected directly to one another.
\end{itemize}

Further details on the experimental hardware can be found in Appendix~\ref{app:experiment-hardware}.

\subsection{Jamming Attack}\label{sec:experiment-design-jamming}

We look first at jamming attacks, using gradient descent to find an optimized jamming signal.
This moves beyond previous work looking at general Gaussian and tone jamming techniques on fingerprinting systems, instead optimizing for maximum disruption to the transmitter fingerprint~\cite{smailesSticky2024}.
This experiment is performed offline, using a trained fingerprinting model and captured data from Iridium satellites, with the channel emulator in the loop to provide realistic attack conditions.

Our goal is to find a generalized jamming signal that works across many different messages which, when added to the synchronization header for a victim message, increases the distance from the message fingerprint to a reference example so that the message is rejected.
To find this, we perform gradient descent directly on the samples of the jamming signal.
At each step of gradient descent, we filter the jamming signal, normalize it to a fixed attacker-to-victim power ratio, and add it to the victim's signal.
The loss function is configured to maximize the distance between the fingerprint of the jammed signal and the fingerprint of the original victim signal.

To find a jamming signal that can be generalized across messages, we apply the jamming to a set of messages during training, taking the mean fingerprint distance of the resulting jammed signals when calculating loss.
We also optionally rotate the jamming signal to a random phase offset after each use, forcing the gradient descent to produce a jamming signal that can work at any phase offset, and is therefore effective even when the attacker cannot achieve phase synchronization.
Finally, we filter the jamming signal to prevent unrealistic wideband interference, using a Finite Impulse Response (FIR) filter with a cutoff of \qty{0.333}{\mega\hertz} and \num{128}~taps.

We control the following parameters:
\begin{itemize}[noitemsep]
    \item Attacker-to-victim power ratio (\mathqty{[-75, 5]}{\decibel})\footnote{At the higher end of this scale, messages fail to decode before reaching the fingerprinter. However, our analysis does not rely upon the decoder, so we can assess the effect on the fingerprinting system independently.}
    \item Number of messages used during training ($1, 10, 100$)
    \item Attacker phase synchronization (True, False)
    \item Channel model (Best Case, Worst Case, Wired Channel)
\end{itemize}

To assess the effectiveness of a jamming signal, we apply it to a new set of clean messages, unseen during the original gradient descent.
We set the acceptance threshold (i.e., the distance between two fingerprints below which the message is accepted) such that \qty{95}{\percent} of legitimate messages are accepted,\footnote{Lowering this threshold accepts more legitimate messages, but enables easier spoofing (and vice versa). \qty{95}{\percent} was chosen as a good middle ground.} and look at the resulting False Rejection Rate (FRR) as the power of the jamming signal increases.
We compare this against prior work's evaluation of Gaussian jamming~\cite{smailesSticky2024}, in which it was shown that an attacker-to-victim power ratio of \qty{-3.0}{\decibel} was sufficient to cause a \qty{50}{\percent} error rate against the Iridium decoder, and \qty{-2.7}{\decibel} against the fingerprinting system.

\subsection{Data Poisoning Attack}\label{sec:experiment-design-poisoning}

Next we evaluate the effectiveness of poisoning attacks, in which stored example messages are gradually replaced by adversarial examples, so that the attacker's messages are accepted as legitimate.
In this attack, we assume the attacker has access to the weights of the fingerprinting model and the example messages stored by the fingerprinter.\footnote{Similar results can be achieved by training a surrogate model on the original dataset to mirror the behavior of the real-world deployment~\cite{demontisWhy2019}.}
As discussed in Section~\ref{sec:threat-model}, we also assume the attacker has some mechanism by which they can introduce their adversarial samples into the fingerprinting system.

The fingerprinting system at the receiver stores an example message for each transmitter, and accepts incoming messages when the distance in fingerprint space between the message and its example falls below a given threshold $a \geq 0$.
Example messages are replaced when the fingerprint falls below the acceptance threshold but above a separate update threshold $u \geq 0$ ($u < a$).
We use gradient descent to construct a short sequence of messages such that no message is rejected by the fingerprinter, triggering updates so that the attacker's messages are accepted by the end of the sequence.
This mirrors the behavior of previous attacks on biometric systems~\cite{biggioAdversarial2015}, and generating adversarial inputs that fall between these two thresholds is a key challenge in poisoning attacks.
If the system uses a different update condition, the attack can be trivially modified to work with the new condition.

Although autoencoders can permit simple interpolation, the specific architecture of \sysname{} raises some issues with this technique; the decoder and encoder are not perfect inverses of one another, so interpolation can result in waveforms whose fingerprints differ quite significantly from one another, particularly at the start and end of the chain.
We instead find that attacks are more successful when only the encoder is used, and the next step is found by gradient descent on the raw samples in the waveform.
At each step $i$, we perform gradient descent starting from the previous message $S_i$, looking for a new message header $S_{i+1}$ which satisfies the following conditions:
\begin{gather*}
u < \text{dist}(e(S_i), e(S_{i+1})) < a \\
\text{dist}(e(S_{i}), e(S_N)) - \text{dist}(e(S_{i+1}), e(S_N)) > u
\end{gather*}
Where $e$ is the fingerprint encoding function, $\text{dist}$ is the distance function, $a$ and $u$ are the acceptance and update thresholds for the fingerprints, and $S_N$ is the target.
A message header which satisfies these conditions will, at each step, be accepted by the fingerprinter and trigger an update to the example message.
These conditions are used as the exit condition for the gradient descent, and the loss function has two corresponding components: one to encourage $\text{dist}(e(S_i), e(S_{i+1}))$ to be between $u$ and $a$, and another to encourage $\text{dist}(e(S_{i+1}), e(S_N))$ to be as small as possible.
This enables us to iteratively move the fingerprint closer to the target, until the final step at which the target's messages are accepted.

The following update condition (with corresponding loss component) can also be added to the process:
\begin{gather*}
\text{dist}(e(S_{i+1}), e(S_0)) < a
\end{gather*}
This results in an ``inclusive poisoning'' attack, in which both the attacker and victim transmitter are accepted as legitimate.
Although the attack is harder to execute due to stricter update conditions, the result is more subtle, since the original victim transmitter is no longer rejected by the fingerprinting system, in addition to the attacker's messages getting accepted.
The attack is therefore more likely to persist for longer without being corrected.

In these experiments we control the following variables:
\begin{itemize}[noitemsep]
    \item Fingerprinter acceptance threshold ($a \in [0, 1]$)
    \item Fingerprinter update threshold ($u \in [0, 1], u < a$)
    \item Inclusive fingerprinting update condition (True, False)
\end{itemize}

We assess the effectiveness of poisoning by looking at how many steps it takes to get between two messages.
We consider the attack to have failed if it takes more than \num{50}~steps, or if the gradient descent does not find a suitable next step within \num{1000}~iterations.

\subsection{Optimized Spoofing Attack}\label{sec:experiment-design-fingerprint-masking}

Finally, we test a spoofing scenario in which the attacker is trying to impersonate a specific transmitter.
To achieve this, they must counteract the effect of their own SDR on the fingerprint, such that it cannot be distinguished from legitimate messages by the receiver.

\subsubsection{Gradient Descent}

Within this scenario, we look first at a simple attack based on gradient descent.
Similar to the jamming attacks, gradient descent is performed on the samples of the spoofing signal added to messages.
However, instead of using legitimate messages, we instead use messages which have already been passed through the physical transmit-receive loop and channel model.
These, with the attacker's additions, are then compared to legitimate messages from the same transmitter, to see if the attacker can successfully remove the fingerprint of their own hardware.

\subsubsection{GAN}

\begin{figure}
    \centering\includegraphics[width=.99\linewidth]{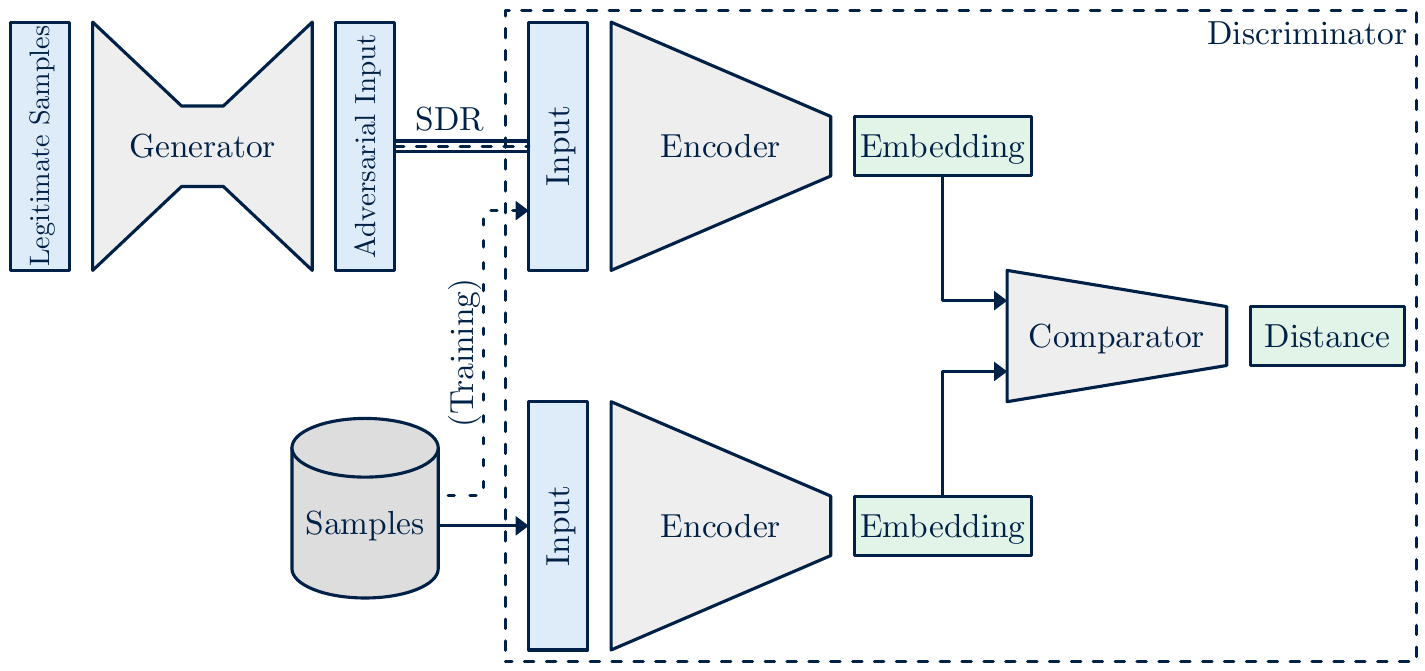}
    \caption{Architecture of the ``Siamese GAN'' model. Legitimate messages are modified by the generator to counteract the fingerprint of their transmitter and channel effects (``SDR''). The discriminator compares these to legitimate messages. During training, pairs of legitimate messages are also used for comparison.}
    \label{fig:gan-siamese}
\end{figure}

Following this, we move on to a GAN architecture, resulting in a system which can counteract impairments in a more general sense.
The architecture of this model is illustrated in Figure~\ref{fig:gan-siamese} (full layers in Appendix~\ref{app:gan-layers}): a generator is trained to create convincing fake messages, at the same time that a discriminator is trained to distinguish them from legitimate messages.
All generated messages are passed through the hardware transmit-receive loop and channel model, modifying the fingerprint of the message via realistic physical conditions in addition to the attacker's transmitter.
This forces the generator to learn how to remove these impairments and convincingly spoof messages, at the same time that the discriminator is learning to detect them.

One beneficial side effect of this architecture is that training does not require the original model, and neither is the resulting GAN dependent on it.
Furthermore, the trained discriminator may also be used as a fingerprinting system itself, since it has been trained directly on identifying optimized spoofing attacks; we evaluate the effectiveness of this fingerprinter in Section~\ref{sec:single-transmitter-fingerprinting}.

Once a model has been trained, we measure its performance by testing on a new set of messages not present in the training data.\footnote{Since our focus is on spoofing attacks, we look at pairs of messages belonging to the same transmitter as each other, one of which has been passed through the GAN.}
We also look at the effect of the transmit-receive loop only (without the corrections added by the GAN), and compare the effectiveness of the GAN's discriminator to the original fingerprinting model.
Finally, we test the spoofing attack using new SDRs not seen during training, in order to find out if the attack transfers between SDRs without retraining and to see how well the GAN works as a countermeasure.

For this experiment, we only need to control the channel model configuration, using the same options as used previously (best case, worst case, wired channel).

%% file: results.tex
\section{Results}\label{sec:results}

\begin{figure}
    \centering\includegraphics[width=.99\linewidth]{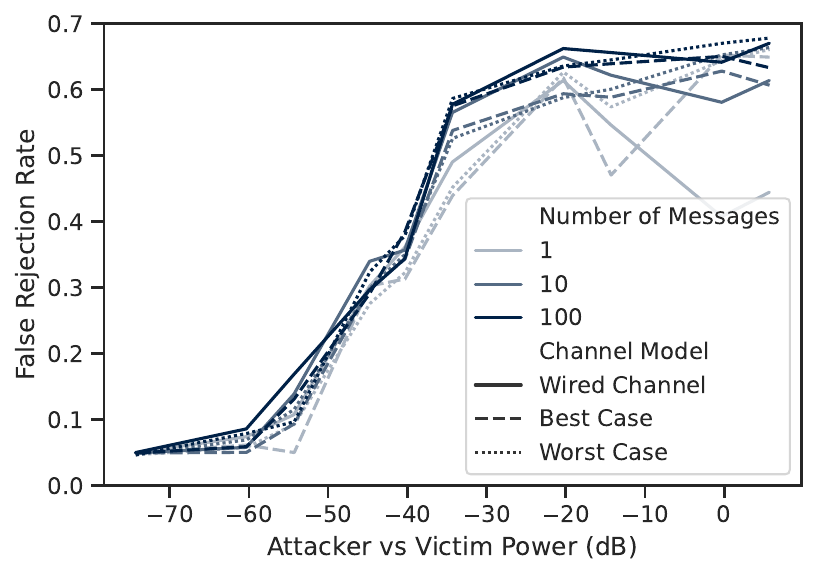}
    \caption{Results from the jamming experiments, when the attacker does not have phase synchronization. False rejections can be induced even at low amplitudes, and the channel model does not have a significant impact on the results.}
    \label{fig:jamming-gradient-descent-combined}
\end{figure}

In this section we implement each of the attacks described above and assess their effectiveness.
We demonstrate that optimized jamming attacks can cause effective denial of service even at low amplitudes, and show how poisoning attacks can gradually shift stored examples so that the system accepts an attacker-controlled transmitter.
We also evaluate spoofing attacks under simple gradient descent and a GAN architecture.
Finally, we evaluate a new single-transmitter fingerprinting technique arising from the trained GAN, and show that it can detect spoofing attacks from previously unseen transmitters.
We focus on a representative selection of results in this section; any remaining results can be found in Appendix~\ref{app:extended-results}.

\begin{figure}
    \centering\includegraphics[width=\linewidth]{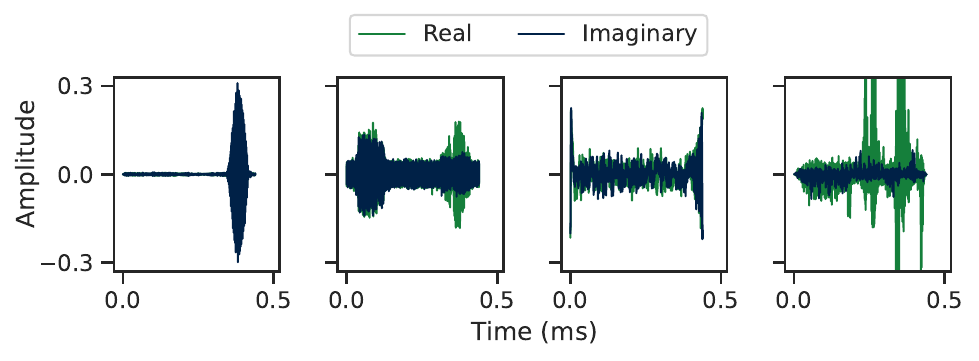}
    \includegraphics[width=\linewidth]{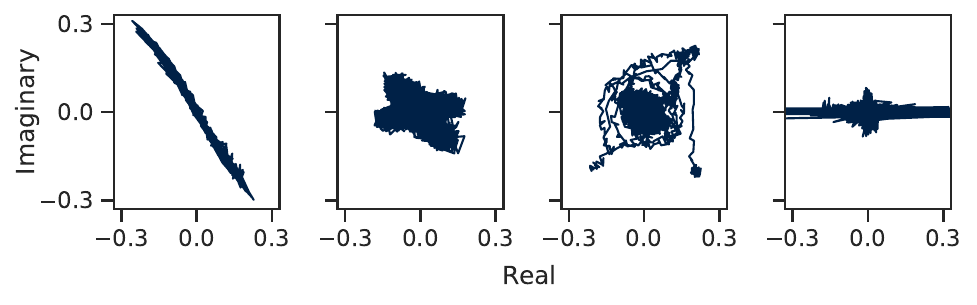}
    \caption{Examples of some of the signals produced by the jamming attack, in the time domain (upper) and as constellation plots (lower).}
    \label{fig:jamming-examples}
\end{figure}

\subsection{Jamming}\label{sec:results-jamming}

\begin{figure*}
    \hspace*{\fill}
    \begin{subfigure}{.45\linewidth}
        \centering\includegraphics[width=\linewidth]{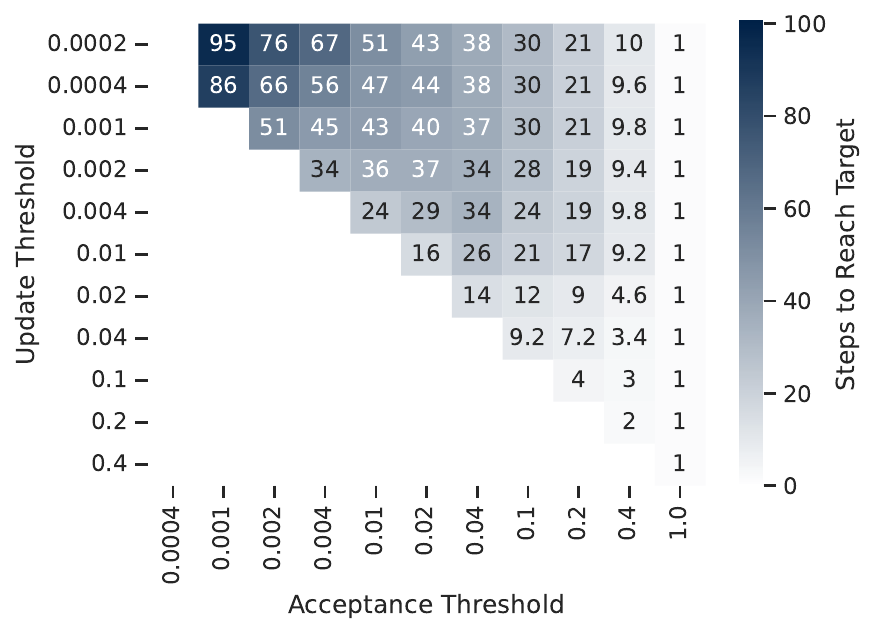}
        \caption{``Exclusive'' poisoning: original transmitter not included in the target.}
        \label{fig:poisoning-heatmap-exclusive}
    \end{subfigure}
    \hfill
    \begin{subfigure}{.45\linewidth}
        \centering\includegraphics[width=\linewidth]{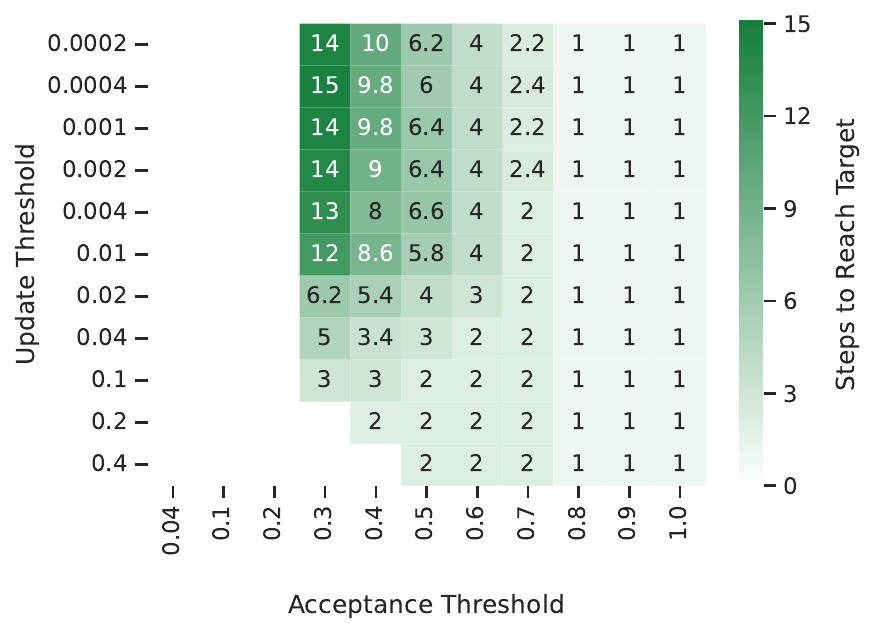}
        \caption{``Inclusive'' poisoning: original transmitter included in the target.}
        \label{fig:poisoning-heatmap-inclusive}
    \end{subfigure}
    \hspace*{\fill}
    \caption{Number of steps required to poison the reference fingerprints to accept one transmitter instead of another, given the acceptance threshold and update threshold. Inclusive poisoning is significantly more difficult and fails for stricter acceptance thresholds, so a different scale is shown.}
    \label{fig:poisoning-heatmap}
\end{figure*}

We start by looking at the results from the jamming attacks described in Section~\ref{sec:experiment-design-jamming}; these are summarized in Figure~\ref{fig:jamming-gradient-descent-combined}.\footnote{Results when the attacker can achieve phase synchronization with the victim can be found in Appendix~\ref{app:extended-results-jamming}; performance is very similar.}
We can see that even at very low amplitudes it is easy to generate adversarial perturbations that disrupt the victim's fingerprint, and that by training on a larger number of example messages, performance is improved slightly on the previously unseen testing messages.
In all cases, the jammer significantly exceeds the performance of traditional Gaussian jamming techniques evaluated in prior works, which demonstrate a \qty{50}{\percent} error rate at approximately \qty{-3.0}{\decibel}~\cite{smailesSticky2024}.
In contrast, we can see that our optimized jamming signal only requires approximately \qtyrange[range-units=single]{-40}{-30}{\decibel} -- this is well below the noise floor, making the attack very difficult to detect.
In practice, the attacker may wish to increase the amplitude of their jamming signal higher than strictly necessary in order to ensure effectiveness in a noisy environment with high levels of attenuation, particularly when they lack any feedback regarding the success of the attack.

Alongside its effectiveness this technique is also versatile, requiring no real-time input of the victim signal or generative capabilities to disrupt communication.
Real-time reactive jamming has been shown to be technically feasible~\cite{moserDigital2019}, but it is difficult and requires high-end hardware; it is much easier to prepare a jamming signal in advance and broadcast it in time with the message header.
We can see from the results that the characteristics of optimized jamming signals are largely message- and transmitter-agnostic, so the attacker does not need to capture the message in real time, and can instead cause significant disruption by using a generalized jamming signal.

Looking at the actual jamming signals produced by this technique, they appear to converge upon a number of different jamming modes, illustrated in Figure~\ref{fig:jamming-examples} (left to right):
\begin{itemize}
    \item High-amplitude bursts, aligned with symbol transitions;
    \item Jamming resembling a QPSK-modulated signal;
    \item Swirl patterns resembling tone jamming;
    \item Sharp bursts aligned with I and Q axes (only emerges when phase alignment is enabled).
\end{itemize}
These are surprisingly similar to known jamming techniques against the decoder~\cite{amuruOptimal2015}; it is possible that both could be deployed simultaneously for greater effect.
We also note that the short impulses used in the final jamming mode may be ineffective in practice, due to the presence of filters on the receiver.
They do, however, help confirm the theory that fingerprint information is derived from the transitions between symbols, as the jamming impulses are aligned with these sections.

\subsection{Poisoning}\label{sec:results-poisoning}

We look next at the poisoning attacks described in Section~\ref{sec:experiment-design-poisoning}, in which the fingerprinter's stored examples are gradually updated to include the fingerprint of an attacker-controlled SDR.
We demonstrate this technique on legitimate messages from the training dataset, poisoning the fingerprinter such that it recognizes messages from transmitter $B$ as if it were transmitter $A$.
An example of the steps involved in interpolating between two fingerprints can be found in Appendix~\ref{app:extended-results-poisoning}.
In order to produce properly optimized signals, the attacker will need to obtain the approximate parameters of the receiver, and adjust the software filtering so the generated messages properly match.

\begin{figure}
    \centering\includegraphics[width=.95\linewidth]{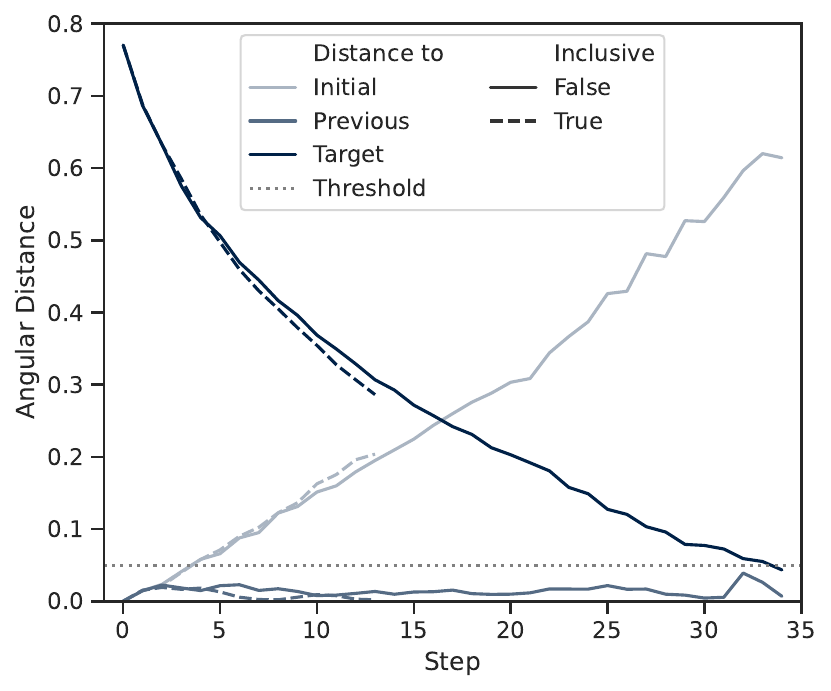}
    \caption{The distance in fingerprint space at each step of the poisoning process, to the initial and target message. Exclusive poisoning succeeds, but inclusive poisoning is forced to terminate early when it fails to find a next step.}
    \label{fig:poisoning-distance}
\end{figure}

This poisoning technique can be repeated for any acceptance threshold and target threshold; in Figure~\ref{fig:poisoning-heatmap} we show the number of steps required for a range of thresholds.\footnote{To ensure computation finished in a timely manner, gradient descent was aborted after \num{1000}~iterations if a suitable next step was not found.}
As the acceptance threshold decreases poisoning becomes increasingly difficult, taking more and more steps and eventually failing to converge for the lowest values.
This effect is more pronounced in the more difficult case of inclusive poisoning, and with thresholds below approximately \num{0.3} the technique fails to find a suitable next step.
Below this value, the fingerprints of the two transmitters are sufficiently far apart that it is not possible to find a fingerprint that includes both the victim's original transmitter and the attacker's transmitter, making inclusive poisoning impossible.

Figure~\ref{fig:poisoning-distance} shows how the distance in fingerprint space changes over time during a poisoning attack, for both exclusive and inclusive fingerprinting.
In both cases, the distance between subsequent steps remains low thanks to the conditions imposed on the gradient descent process, and the distance to the target fingerprint gradually decreases until it falls below the acceptance threshold, at which point the attack is considered successful.
In the inclusive case, the strategy is forced to terminate earlier as the system fails to find a suitable next step.
Despite this, an attacker may be able to use inclusive poisoning when the acceptance threshold is sufficiently high, and in other cases may take advantage of inclusive poisoning to evade detection for longer -- the initial stages of poisoning can be completed inclusively to evade detection, switching only to exclusive poisoning when forced to do so.

\begin{figure}
    \centering\includegraphics[width=.95\linewidth]{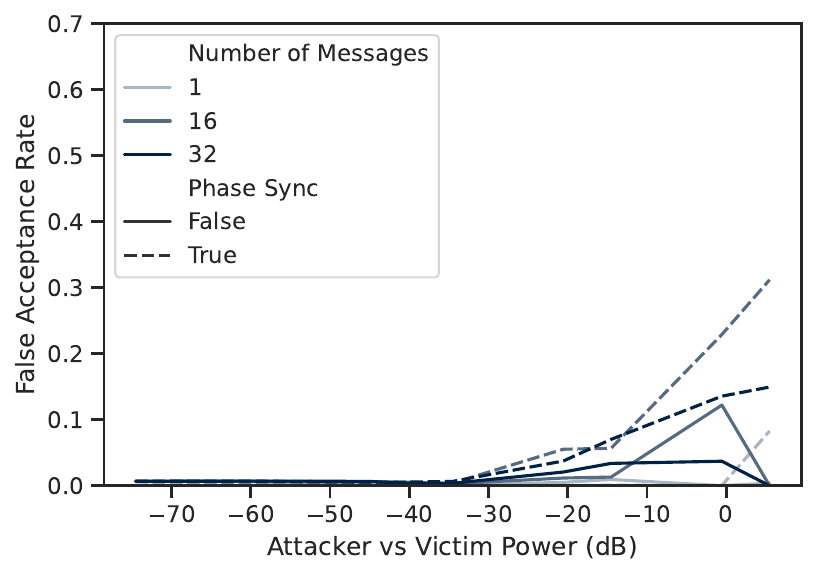}
    \caption{Results from the spoofing attack using simple gradient descent to find an optimized signal. Performance is poor even when phase synchronization is permitted; a more advanced technique is required.}
    \label{fig:spoofing-gradient-descent-worstcase}
\end{figure}

We also carried out the poisoning attack on a random fingerprint, showing that the system can still find a sequence of messages which result in this random fingerprint being recognized as legitimate.
An example of this process can also be found in Appendix~\ref{app:extended-results-poisoning}, taking a similar number of steps to the original poisoning attack.
We can therefore be confident that this attack will work even on transmitters outside the original dataset.

Through these results we have shown that the fingerprinter can be poisoned to accept any target transmitter as legitimate, by injecting a short sequence of controlled messages to guide the reference examples away from the original transmitter.
This can certainly be achieved using an AWG, but it remains to be seen whether off-the-shelf SDR hardware could deliver the same result.
In Section~\ref{sec:discussion} we discuss improvements to the fingerprint update function which might mitigate this attack.

\subsection{Optimized Spoofing}\label{sec:results-fingerprint-masking}\label{sec:results-optimized-spoofing}

Finally we look at the results for the fingerprint masking attacks, starting with simple gradient descent before moving to the GAN-based attacks.

\subsubsection{Gradient Descent}

First we look at the simple masking attack, using the same architecture as in the jamming attacks but using SDR-replayed messages, using the experimental setup from Section~\ref{sec:hardware-configuration}.
Results from this attack are summarized in Figure~\ref{fig:spoofing-gradient-descent-worstcase}, for the ``Worst Case'' channel model configurations.%
\footnote{The results for the other configurations are very similar, and can be found in Appendix~\ref{app:extended-results-fingerprint-masking}.}
It is clear from these results that simple gradient descent as a technique is insufficient to figure out how to counteract the attacker's fingerprint -- even when phase synchronization is permitted, false acceptance rates remain below \qty{10}{\percent} until the modifications become so powerful that the attacker can simply replace the signal entirely, and even then barely exceed \qty{30}{\percent}.

In addition to its limited success rate, this approach is of limited utility in a real-world setting, since the attacker makes their modifications after the message has already been transmitted over the air.
For the attack to be more effective, the model must learn to counteract the fingerprint of their transmitter and other physical layer effects \textit{before} the message is transmitted.

\subsubsection{GAN}\label{sec:results-gan}

\begin{figure*}
    \centering\includegraphics[width=.95\linewidth]{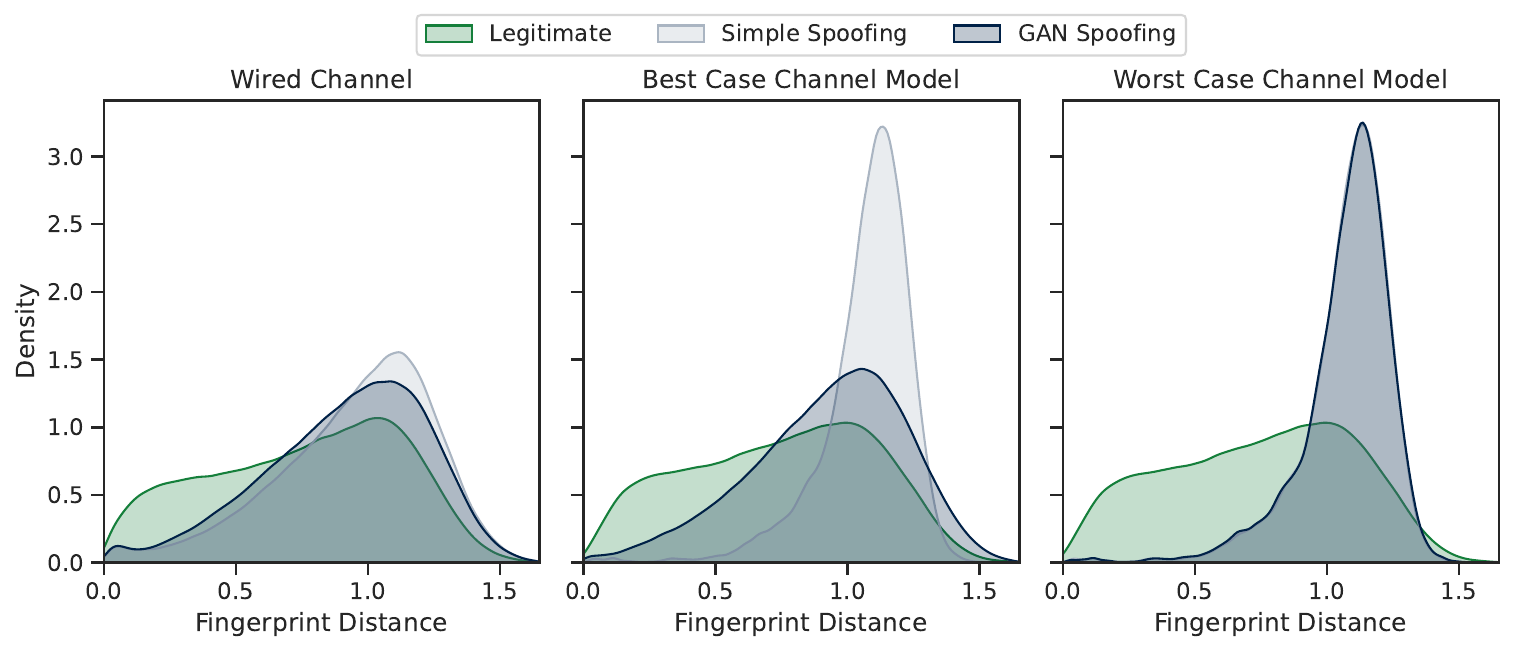}
    \caption{Distribution of distances in fingerprint space for each attack between pairs of legitimate messages, legitimate and replayed messages (``simple spoofing''), and legitimate and replayed messages using the GAN (``GAN spoofing''). In all cases except for the worst case channel model, the GAN is able to counteract some of the effects of replaying messages for a more successful attack.}
    \label{fig:histograms-spoofing-satiq}
\end{figure*}

We therefore look next at the GAN architecture, which uses the real-time SDR loop as feedback during training, and tweaks messages to counteract physical layer effects with maximal effectiveness.
We assess performance of the attack by looking at the distribution in fingerprint space of pairs of messages, given in Figure~\ref{fig:histograms-spoofing-satiq}.
We can also assess the overall performance by looking at the Receiver Operating Characteristic (ROC) curve, which plots the true positive rate against the false positive rate as the acceptance threshold changes.
The ROC Area Under Curve (AUC) indicates the overall performance across all thresholds, with ideal performance at \num{1.0} and random guessing at \num{0.5}.
We also look at the Equal Error Rate (EER): the point at which the false accept and false reject rates are the same.
These are given in Table~\ref{tab:results-satiq}.

\begin{table}
    \caption{Performance of simple replay attacks and the GAN-optimized spoofing attack against the \sysname{} fingerprinting model, for each of the three channel models. In all but the worst channel conditions, the attacker is able to counteract SDR and channel effects.}
    \label{tab:results-satiq}
    \autobox{
    \begin{tabular}{llS[table-format=1.4]S[table-format=1.4]}
        \toprule
        Channel Model & Spoofing & {AUC} & {EER} \\
        \midrule
        Wired Channel & Simple & 0.6553 & 0.3909 \\
                      & GAN & 0.6224 & 0.4191 \\
        Best Case & Simple & 0.7887 & 0.2643 \\
                  & GAN & 0.6482 & 0.3979 \\
        Worst Case & Simple & 0.7894 & 0.2635 \\
                   & GAN & 0.7877 & 0.2654 \\
        \bottomrule
    \end{tabular}
    }
\end{table}

From these results, we can see the difference between legitimate and replayed messages (``simple spoofing''), and the effect of the GAN in removing the effect of the spoofing attack.
When the wired channel is used, the spoofing attack is moderately effective even without the GAN thanks to the cleanness of the signal, although the GAN is able to partially remove the small differences to the fingerprint, for an EER of \qty{41.9}{\percent}.
For the ``best case'' wireless channel, replayed messages look substantially different from legitimate messages, but the attacker's model is able to counteract these changes quite well.
In the ``worst case'' channel, the fading effects from the channel dominate and the GAN struggles to consistently undo them.

To test the attack's transferability, we also tested the attack using the same transmit-receive loop as before, but with different SDR frontends.
The results from this analysis are in Figure~\ref{fig:histogram-spoofing-satiq-bestcase-b} for the ``best case'' channel model (remaining results in Appendix~\ref{app:extended-results-gan}), showing almost identical performance as the original attack -- it is therefore likely that the signal characteristics the GAN has learned to remove are consistent between instances of the same SDR hardware and channel conditions, enabling the attacker to train a single model and carry out the same attack using multiple SDRs.

\subsection{Single-Transmitter Fingerprinting}\label{sec:single-transmitter-fingerprinting}

Although the primary goal of training a GAN was to carry out spoofing attacks against the fingerprinter, the nature of the model means it also necessarily results in the creation of a discriminator for detecting these spoofing attacks.
Unlike the \sysname{} model, which was trained on identifying transmitters and gains spoofing/replay detection incidentally, these discriminators are explicitly trained to detect attacks.
We therefore test it against replayed messages from three sources:
\begin{itemize}[noitemsep]
    \item The original transmit-receive loop and channel model used during training;
    \item A modified transmit-receive loop with the SDRs exchanged, to modify the attacker's fingerprint;
    \item The replay dataset provided by the authors of \sysname{}~\cite{smailesWatch2023}.
\end{itemize}

\begin{figure}
    \centering\includegraphics[width=.85\linewidth]{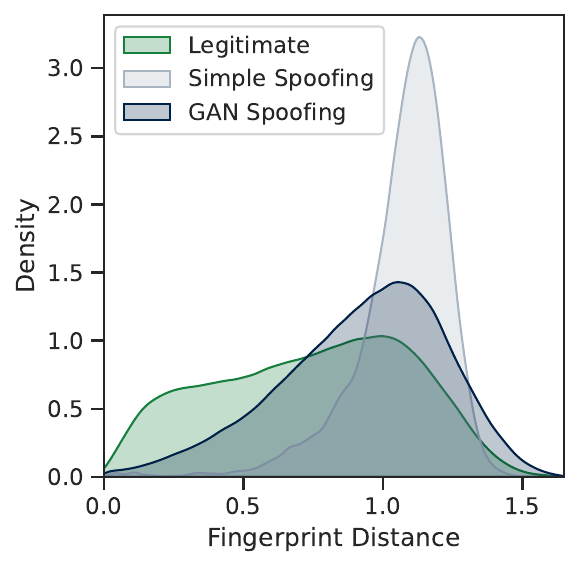}
    \caption{Distribution of distances in fingerprint space when the attacker is using new SDR hardware not used during training, for the ``best case'' channel conditions. Results are almost identical to those from the training hardware.}
    \label{fig:histogram-spoofing-satiq-bestcase-b}
\end{figure}

\begin{figure}
    \centering\includegraphics[width=.95\linewidth]{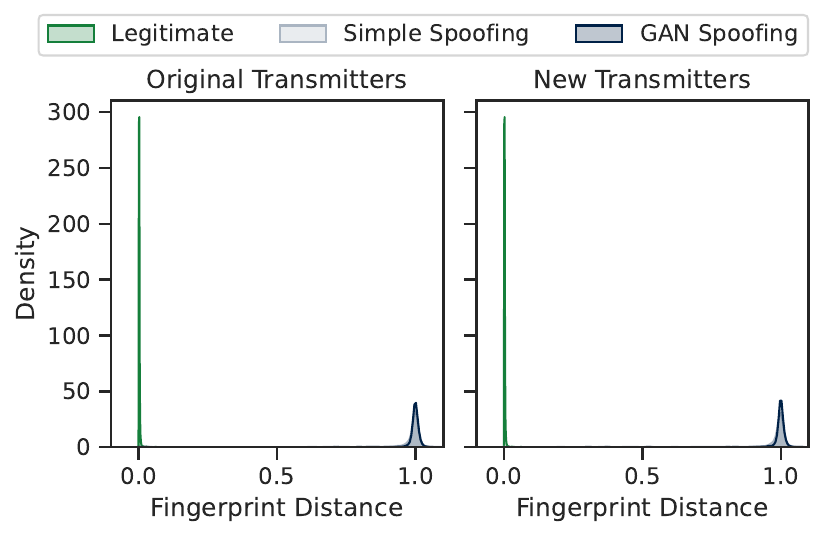}
    \caption{Distribution of distances in fingerprint space using the newly trained GAN, for replay attacks using the SDRs used in training (left), and previously unseen SDRs with the same hardware (right). The model can easily identify replay attacks with near certainty when using similar hardware to that seen in training.}
    \label{fig:histograms-single-transmitter}
\end{figure}

\begin{figure*}
    \begin{subfigure}{.498\linewidth}
        \centering\includegraphics[width=.95\linewidth]{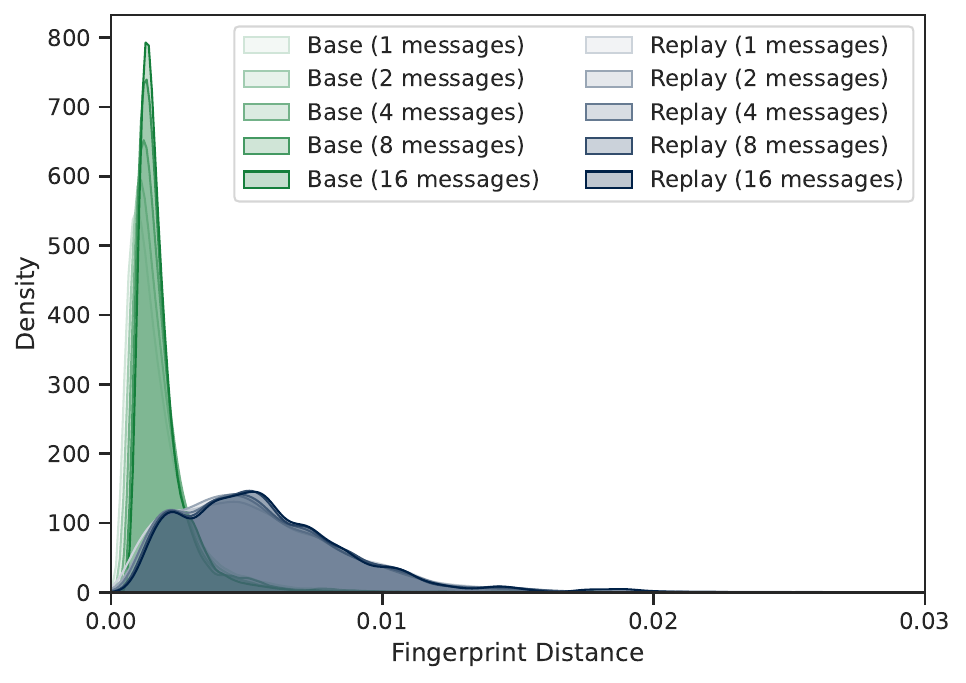}
        \caption{GAN discriminator.}
    \end{subfigure}
    \begin{subfigure}{.498\linewidth}
        \centering\includegraphics[width=.95\linewidth]{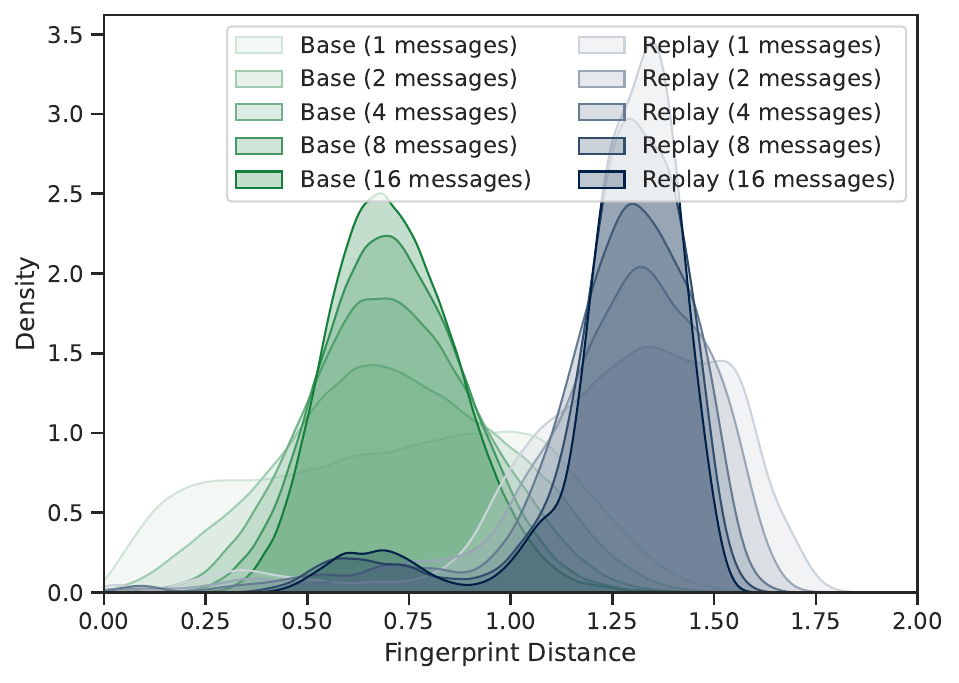}
        \caption{\sysname{} model.}
    \end{subfigure}
    \caption{Distances in fingerprint space when detecting replay attacks from the \sysname{} replay dataset, tested against the new GAN and the original \sysname{} model. Performance is improved by comparing each incoming message against multiple examples.}
    \label{fig:histograms-satiq-multiple-examples}
\end{figure*}

The results of this analysis on the transmit-receive loop are shown in Figure~\ref{fig:histograms-single-transmitter} and Table~\ref{tab:results-single-transmitter}, for ``worst case'' channel conditions (``best case'' results in Appendix~\ref{app:extended-results-single-transmitter}).
We can see that it is very easy for the discriminator to separate legitimate messages from SDR-replayed ones, even when the hardware has been exchanged, achieving an EER of \qty{0.4}{\percent}.
This suggests that there are clear characteristics unique to the attacker's SDR which can be detected with explicit training, but that the \sysname{} model struggles to detect, due to having only been trained on legitimate Iridium transmitters.
In Figure~\ref{fig:histograms-satiq-multiple-examples} and Table~\ref{tab:results-single-transmitter-satiq} we evaluate the GAN against the \sysname{} replay dataset.
In this dataset, the fingerprints are much closer to one another but still reasonably separable, with an EER of \qty{15.9}{\percent} when comparing against \num{16} example messages.\footnote{We found that the model trained against the ``worst case'' channel model had the best performance in this area.}
This is slightly worse than the original \sysname{} model's EER of \qty{7.7}{\percent}, but promising considering the model has never seen the attacker's model of SDR in the training data.
We discuss the implications of this result in Section~\ref{sec:discussion}, suggesting how this can be deployed in practice at this level of accuracy, and how performance might be improved by training on a variety of different SDRs.

Since the GAN has been trained directly to detect spoofing attacks, it demonstrates much better performance in this area.
This also opens up some interesting new possibilities in the realm of single-transmitter fingerprinting -- unlike the original \sysname{} model, which requires a large dataset comprising multiple transmitters to train, this architecture needs only a small number of messages from a single transmitter, and a transmit-receive loop to train the system to distinguish between legitimate and replayed messages.
This lies in contrast to all previous works, which focus on larger constellations with many transmitters, and cannot be used to secure individual satellites or transmitters.
A ``conditional GAN'' architecture~\cite{mirzaConditional2014} could also be used to bridge the gap to providing transmitter-specific authentication in a small constellation, given a sufficiently large dataset of legitimate messages from each transmitter.
With recent increases in attacks on satellite systems, even those equipped with cryptographic security, any additional authenticity and signal intelligence that can be applied on top of existing systems is invaluable.
By expanding the scope of previous works, our results enable the detection and mitigation of attacks even on individual satellites, thereby providing more robust and resilient communication and enhancing the overall security of satellite-dependent infrastructure.

\begin{table}
    \caption{AUC and EER values from the results in Figure~\ref{fig:histograms-single-transmitter}.}
    \label{tab:results-single-transmitter}
    \autobox{
    \sisetup{round-mode=places,round-precision=4}
    \begin{tabular}{lS[table-format=1.4]S[table-format=1.4]}
        \toprule
        Replay Attack & {AUC} & {EER} \\
        \midrule
        Original & 0.999626 & 0.003658 \\
        Unseen SDRs & 0.999336 & 0.004192 \\
        \bottomrule
    \end{tabular}
    }
\end{table}

%% file: discussion.tex
\section{Discussion}\label{sec:discussion}

In this paper we have demonstrated the vulnerability of radio frequency fingerprinting to jamming, dataset poisoning, and optimized spoofing attacks, using Iridium as a case study.
While these attacks can be effective, there are some limitations and potential mitigations that can be explored, improvements to the proposed single-transmitter fingerprinting technique, and other interesting avenues of future work.

\begin{table}
    \caption{Performance of the GAN at detecting replay attacks using previously unseen transmitters in the \sysname{} replay dataset, comparing each message against multiple examples. Results are also included for the original \sysname{} fingerprinter for comparison.}
    \label{tab:results-single-transmitter-satiq}
    \autobox{
    \sisetup{round-mode=places,round-precision=4}
    \begin{tabular}{lS[table-format=1.4]S[table-format=1.4]S[table-format=1.4]S[table-format=1.4]}
        \toprule
        Messages & \multicolumn{2}{c}{GAN Discriminator} & \multicolumn{2}{c}{\sysname{} Model} \\
        \cmidrule(lr){2-3} \cmidrule(lr){4-5}
        & {AUC} & {EER} & {AUC} & {EER} \\
        \midrule
        1 & 0.875494 & 0.193684 & 0.889342 & 0.187548 \\
        2 & 0.887999 & 0.176836 & 0.921971 & 0.138952 \\
        4 & 0.893013 & 0.171288 & 0.941920 & 0.096716 \\
        8 & 0.907865 & 0.159168 & 0.952252 & 0.078151 \\
        16 & 0.915557 & 0.158997 & 0.952501 & 0.076901 \\
        \bottomrule
    \end{tabular}
    }
\end{table}

\subsection{Countermeasures}

There are a number of countermeasures which may be explored to increase the difficulty of attacks on fingerprinting systems, or prevent them entirely.
For instance, comparing incoming messages against multiple examples or taking a rolling average from multiple messages can reduce false rejection rate and make attacks more difficult to execute, requiring the attacker to synchronize their signal to multiple messages in a short time period in order to have a significant effect.
The presence of filters in standard receiver hardware may also affect attacks, although our results show that including these filters in the adversarial training pipeline enables attacks to remain effective with a minimal performance drop.

Another approach, providing protection against dataset poisoning, is to prevent attackers from controlling when stored examples are updated.
This could be achieved by triggering updates manually within a given time window, or by randomly choosing from the most recent $N$ messages sent from a given transmitter.
This can make poisoning attacks substantially more difficult to execute, requiring persistent effort from the attacker over an extended period of time.

We can also draw inspiration from existing systems in their approach to countering attacks~\cite{machadoAdversarial2021,adesinaAdversarial2022}.
These include training classifiers on an augmented dataset containing adversarial examples (``adversarial training'')~\cite{jinAPEGAN2019,madryDeep2017}, or using techniques like randomized smoothing and defensive distillation to smooth gradients, thus reducing the attacker's ability to find adversarial perturbations~\cite{papernotDistillation2016,cohenCertified2019,kimChannelAware2021}.
Preprocessing techniques are also proposed, such as training an auxiliary detection model to remove adversarial perturbations prior to the classifier~\cite{gongAdversarial2023,grosseStatistical2017,metzenDetecting2017,kokalj-filipovicTargeted2019} (optionally also taking into account characteristics of the physical RF channel~\cite{kokalj-filipovicMitigation2019}) or using an ensemble of classifiers to counteract one another's weaknesses~\cite{tramerEnsemble2017}.
Finally, ``certified defense'' attempts to verify that a model cannot be attacked by adversarial inputs, providing a certificate that proves that no adversarial perturbation below a given amplitude can result in more than a given level of misclassification on the test data~\cite{raghunathanCertified2018} -- this has been demonstrated on wireless signal classifiers~\cite{kimChannelAware2021}.
Many of these techniques work with distance-based systems, or could easily be adapted to do so, and could therefore result in significantly more robust satellite fingerprinting systems.

Operators may also counter attacks through combining multiple sources of signal intelligence to provide improved coverage against attacks.
Although the optimized attacks in this paper are effective against the fingerprinting system tested, they may be detected by a different mechanism -- for example, the high-amplitude bursts exhibited by some of the jamming signals in Section~\ref{sec:results-jamming} could be identified by monitoring the signal-to-noise ratio over time.
By combining multiple methods, the attack surface of satellite communication systems can be minimized, making them significantly harder to attack.

\subsubsection{Single-Transmitter Fingerprinting}

The single-transmitter fingerprinting technique introduced in this paper offers another promising countermeasure against spoofing and replay attacks -- by training a model directly on detecting attacks, performance is significantly higher than other systems which gain this capability only incidentally, as a side effect of training to distinguish legitimate transmitters.
This is a particularly powerful countermeasure, and can be deployed alongside other methods of detecting and preventing attacks (including other fingerprinting models) without impacting operation.
Our results demonstrate the initial capability of this technique, but the model should be trained against multiple different SDRs prior to deployment in the field, to ensure the system can reliably detect attacks from a wide range of attacker hardware.
The acceptance threshold of the model must also be tuned to minimize false rejections whilst still detecting attacks.
Alternatively, operators may simply use fingerprinting systems as a metric for monitoring and post-attack diagnostic purposes, rather than tying it directly into the underlying authentication system.

Alongside enabling fingerprinting in smaller satellite systems, this technique also opens up the opportunity for centralized training: a single SDR loop system can be used to train many models at once, targeting different satellite systems.
This could even result in system-agnostic fingerprinting, with a model trained to detect attacks regardless of the underlying signal modulation scheme.

\subsection{Future Work}

Our results have shown that fingerprinting techniques are far from immune to attacks, but that they can provide some level of protection, particularly when deployed in a more targeted manner.
Future work should therefore focus on developing and evaluating combined countermeasures to provide broad coverage against various types of attacks.
Additionally, the development of standardized testing frameworks would enable operators to assess their systems' resilience against different types of attacks in a consistent and reliable manner, and deploy new countermeasures.

Future work might also expand the scope of the hardware configuration -- for example, by including a greater number of SDRs during training, or using a larger number of different types of SDR.
Doing so would allow a model to identify attacks in a greater number of settings, granting increased versatility.
It would also be useful to assess the capability of a model trained to detect multiple types of attack (e.g., jamming alongside spoofing), or to combine transmitter identification with attack detection.
It may be the case that these combined models exhibit worse performance than they would if trained separately, in which case multiple models can be used in tandem with one another.

%% file: conclusion.tex
\section{Conclusion}\label{sec:conclusion}

In this paper we have demonstrated a range of optimized attacks against satellite fingerprint authentication.
We demonstrate that optimized jamming attacks can cause false rejections even with very low amplitude perturbations, enabling effective denial of service.
Through dataset poisoning we show that reference messages can be gradually modified such that attacker-controlled transmitters are accepted as legitimate, and through optimized spoofing an attacker can mask out the fingerprint of their attacking SDR to avoid detection.
By demonstrating these threats, we emphasize the need for robust security measures in satellite communication, and the importance of combining multiple sources of signal intelligence in order to achieve broad coverage against attacks.

Alongside this, we have also introduced a new countermeasure to attacks: a GAN trained for optimized spoofing attacks can be repurposed to instead detect spoofing attacks with high performance, even against transmitters it has never seen before.
This advancement also enables single-transmitter fingerprinting -- a technique previously thought to be infeasible due to the requirement of multiple transmitters in the training data.
This represents a significant step forward in enhancing the resilience of satellite systems, offering a vital security improvement to legacy systems that lack cryptographic protection, and providing additional attack detection capabilities for newer systems.

%% file: appendices/experiment-hardware.tex
\section{Experiment Hardware}\label{app:experiment-hardware}

In this appendix we provide further details on the experiment hardware described in Section~\ref{sec:experiment-design}.

Our experiments use the following hardware:
\begin{itemize}[noitemsep]
    \item Ettus Research OctoClock-G CDA-2990
    \item $2\times$ Ettus Research USRP X300 SDR
    \item $2\times$ Ettus Research SBX-120 daughterboard
    \item Mini-Circuits BW-S30W20+ attenuator
    \item Keysight PROPSIM FS-16 channel emulator
\end{itemize}

The first SDR is connected via an SMA cable to the input of the channel emulator.
Its output is connected via another SMA cable to the second SDR, creating a transmit-receive loop in which one SDR can send signals to be received by the other after being subjected to attenuation and fading by the channel emulator.
In the ``wired channel'' case, the SDRs are instead connected directly to one another, with an attenuator inline to prevent damage to the hardware.
The SDRs are synchronized using the OctoClock, providing a common clock signal to ensure they do not drift from one another.

Messages sent through the loop are preceded by a rising edge to enable synchronization at the sample level, and a header with a known phase is used for phase alignment.

On top of this hardware setup we provide a simple ZeroMQ interface.
Any software that sends samples through this interface will receive as a response those same samples after they have been sent through the transmit-receive loop.
This is also incorporated into a TensorFlow layer, allowing the hardware loop to be used for dataset preprocessing or integrated into the model itself, enabling it to learn the characteristics of physical layer distortions and how to counteract them.
All code used in this setup will be released on publication.

%% file: appendices/gan-layers.tex
\section{GAN Layers}\label{app:gan-layers}

\begin{figure*}
    \centering\includegraphics[width=\linewidth]{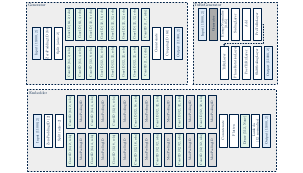}
    \caption{Architecture of the Siamese GAN model, split into its component parts.}
    \label{fig:gan-layers}
\end{figure*}

Figure~\ref{fig:gan-layers} shows the structure of the GAN model used in the fingerprint masking experiments, described in Section~\ref{sec:experiment-design-fingerprint-masking}.
The \textit{Generator} takes message headers and generates modifications designed to remove the fingerprint of the SDR hardware.
The \textit{TxRxGenerator} takes these modifications, adds them to the original message, and passes the result through the transmit-receive loop.
The \textit{Embedder} takes a message header and reduces it to a lower-dimensional fingerprint.
A discriminator is constructed using the embedder, computing the angular distance between a pair of embeddings.

Taking the generator $g$ and discriminator $d$, we then build two loss terms:
\begin{itemize}
    \item For any given input $i$, the generator loss minimizes $d(i, g(i))$, the distance between the input and the input following the generator.
    \item For two inputs $i_a$ and $i_b$, the discriminator loss encourages $d(i_a, g(i_a))$ to be close to $1$ and $d(i_a, i_b)$ to be close to $0$.
\end{itemize}

%% file: appendices/extended-results.tex
\section{Extended Results}\label{app:extended-results}

In this appendix we include the remaining results from our experiments, for the sake of completeness.

\subsection{Jamming with Phase Synchronization}\label{app:extended-results-jamming}

\begin{figure}
    \centering\includegraphics[width=.85\linewidth]{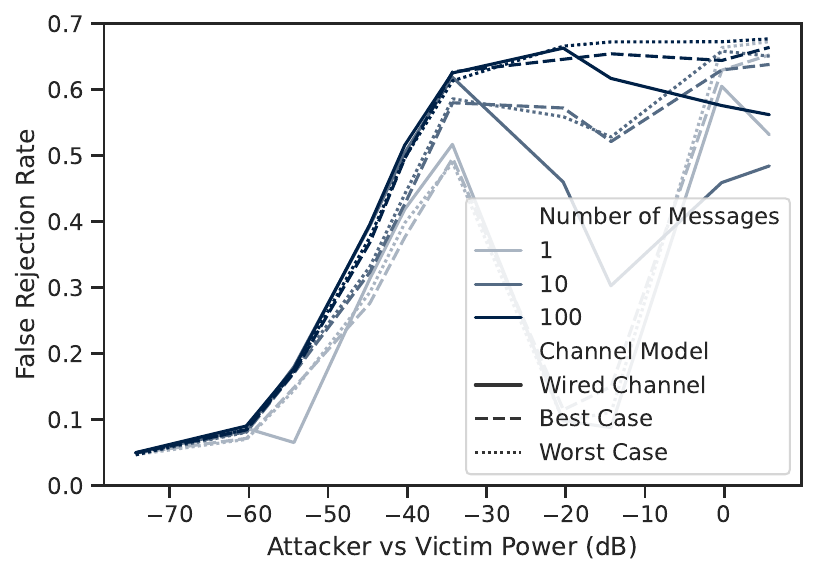}
    \caption{Results (mean fingerprint distance and false rejection rate) from the jamming experiments, when the attacker can achieve phase synchronization with the victim.}
    \label{fig:jamming-gradient-descent-combined-phase}
\end{figure}

Results for the jamming attack when phase synchronization is enabled can be seen in Figure~\ref{fig:jamming-gradient-descent-combined-phase}.
These results are at best comparable to the original results, and in many cases produce worse performance, particularly at higher power levels.
It is likely this is caused by overfitting, with random phase shifts serving to force the jamming signal to be more generalizable -- this means the attacker does not need to worry about phase alignment, and can simply generate a jamming signal and align it at the symbol level.

\subsection{Poisoning}\label{app:extended-results-poisoning}

\begin{figure*}
    \centering\includegraphics[width=\linewidth]{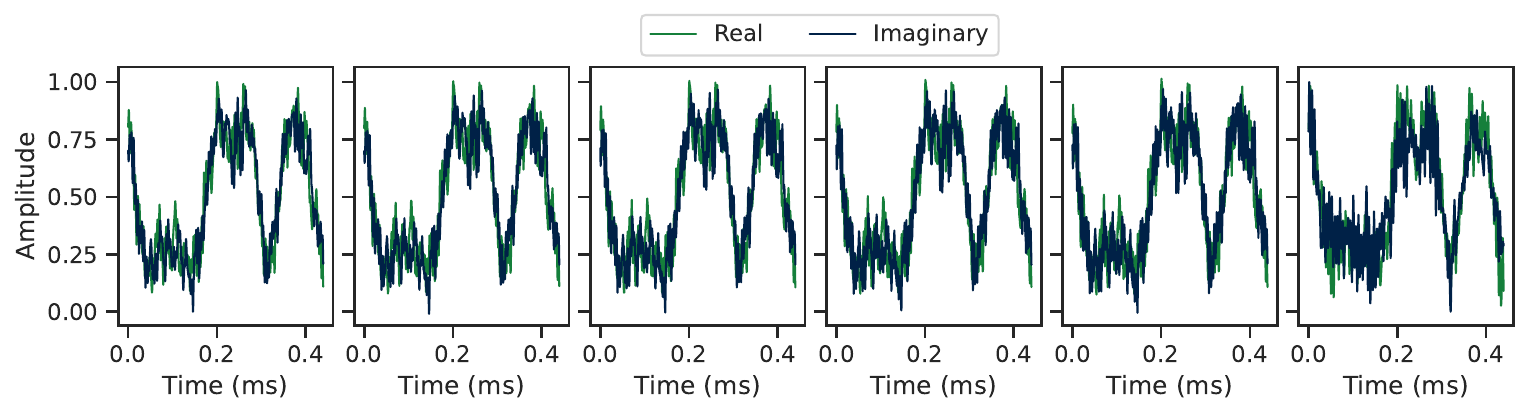}
    \centering\includegraphics[width=\linewidth]{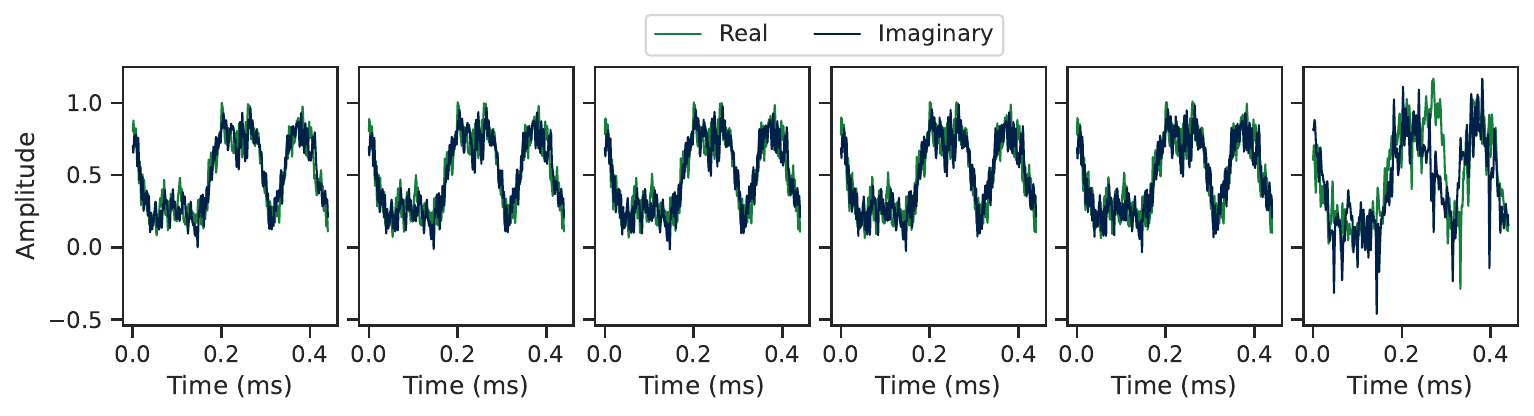}
    \caption{A selection of the steps involved in poisoning by interpolating between the fingerprints of two legitimate transmitters (upper), and between a legitimate transmitter and a random fingerprint (lower).}
    \label{fig:interpolation-samples}
\end{figure*}

Figure~\ref{fig:interpolation-samples} shows a selection of steps in the poisoning process, demonstrating how the message header from one transmitter can gradually change to resemble the message header from another transmitter.
We also show this for a randomly generated message header not present in the dataset, to show that this can be achieved for any type of transmitter.

In these plots we can see the noise-removing effect of the fingerprinter's autoencoder: even though the final waveform generated by the poisoning process has more noise than the target waveform, the fingerprints are still highly similar to one another.
We also see that, similar to the jamming experiments, altering the fingerprint does not require a signal with a huge amplitude.

By demonstraing the poisoning attack on a randomly generated fingerprint, we show that the dataset poisoning attack can work on arbitrary data and fingerprints, even those that do not resemble any legitimate transmitters seen by the model during training.
Note that in this case the final message differs quite noticeably from the generated poisoning samples, likely due to the denoising effect of the fingerprinter.
This works in the attacker's favor, as they can send messages which more closely resemble the fingerprint of the original victim.

\subsection{Optimized Spoofing}\label{app:extended-results-fingerprint-masking}

Finally, we look at the optimized spoofing results summarized in Section~\ref{sec:results-optimized-spoofing}.

\subsubsection{Gradient Descent}\label{app:extended-results-gradient-descent}

\begin{figure*}
    \begin{subfigure}{.495\linewidth}
        \centering\includegraphics[width=\linewidth]{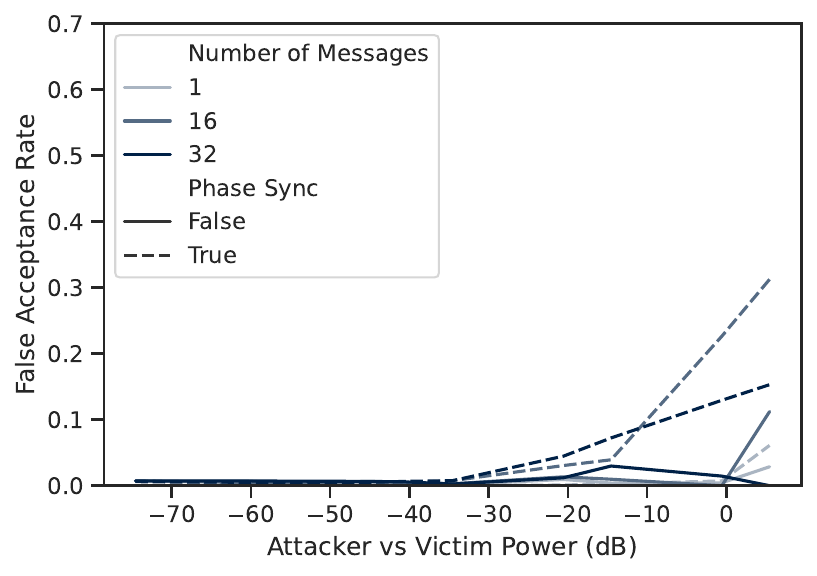}
        \caption{``Best case'' configuration.}
    \end{subfigure}
    \begin{subfigure}{.495\linewidth}
        \centering\includegraphics[width=\linewidth]{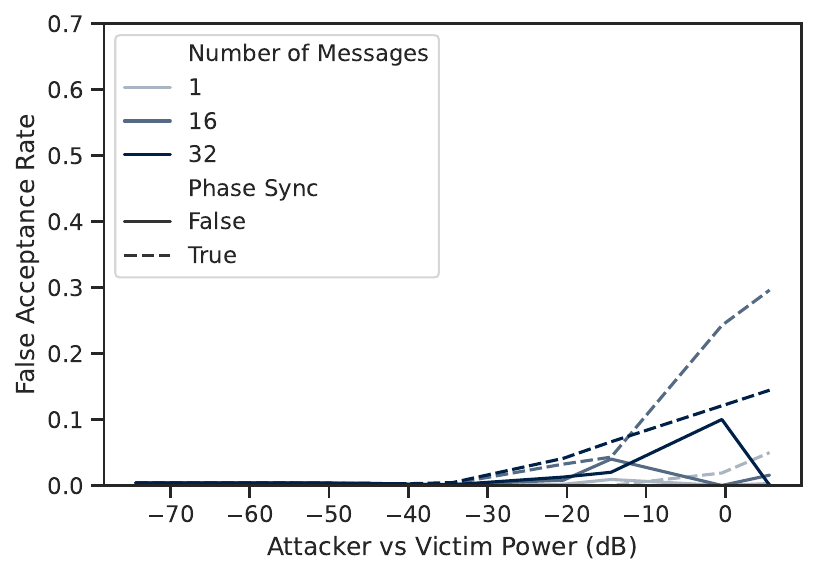}
        \caption{``Wired channel'' configuration.}
    \end{subfigure}
    \caption{Results from the gradient descent spoofing attack for the remaining two channel model configurations. Performance is similar to the results from the ``worst case'' channel model.}
    \label{fig:spoofing-gradient-descent-full}
\end{figure*}

In Figure~\ref{fig:spoofing-gradient-descent-full} we can see the results from the gradient descent spoofing attack when the remaining two channel model configurations are used, building upon the results in Figure~\ref{fig:spoofing-gradient-descent-worstcase}.
We can see that the results are very similar, with the success rate barely exceeding \qty{10}{\percent} when phase synchronization is not permitted, and \qty{30}{\percent} when it is.

\subsubsection{GAN}\label{app:extended-results-gan}

\begin{figure}
    \centering\includegraphics[width=.667\linewidth]{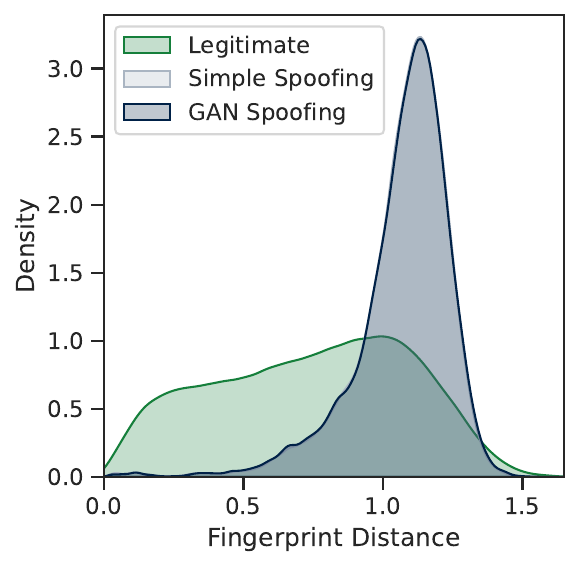}
    \caption{Distribution of distances in fingerprint space when the attacker is using new SDR hardware not used during training, for the ``worst case'' channel conditions. Results are almost identical to those from the training hardware.}
    \label{fig:histogram-spoofing-satiq-worstcase-b}
\end{figure}

In Figure~\ref{fig:histogram-spoofing-satiq-worstcase-b} we see the results from the GAN-based attack using a new SDR frontend not used during training, for the worst case channel model.
Like the results from the best case channel model in Section~\ref{sec:results-gan}, they are nearly indistinguishable from the original results.

\subsection{Single-Transmitter Fingerprinting}\label{app:extended-results-single-transmitter}

\begin{figure}
    \centering\includegraphics[width=.85\linewidth]{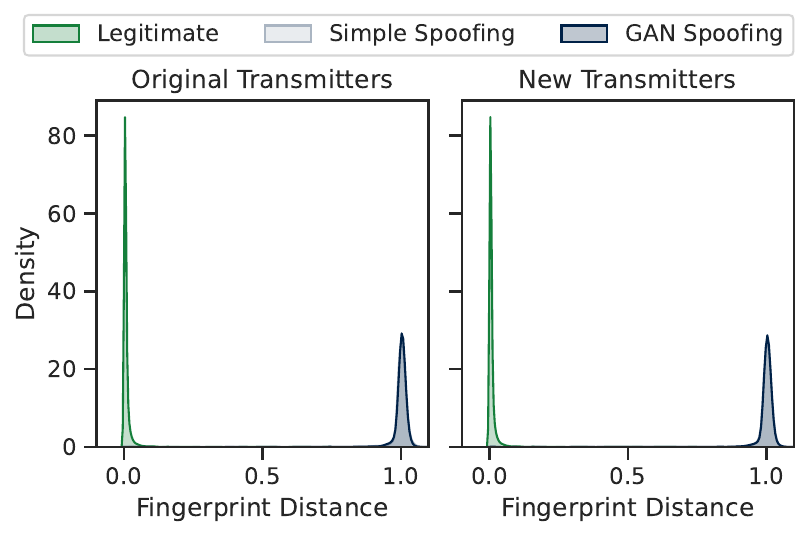}
    \caption{Distribution of distances in fingerprint space using the newly trained GAN, for replay attacks using the SDRs used in training (left), and previously unseen SDRs with the same hardware (right), using the ``best case'' channel model.}
    \label{fig:histograms-single-transmitter-bestcase}
\end{figure}

Figure~\ref{fig:histograms-single-transmitter-bestcase} shows the single-transmitter fingerprinting evaluation from Section~\ref{sec:single-transmitter-fingerprinting} for the ``best case'' channel model.
The results are almost identical, with the model easily able to distinguish the legitimate and replayed messages.